# Achromatic Non-Interferometric Single Grating Neutron Phase Imaging

M. Strobl,[1,2*] J. Valsecchi,[1,$] R.P. Harti,[1] P. Trtik,[1] A. Kaestner,[1] C. Gruenzweig,[1] E. Polatidis,[1] J. Capek[1]

[1]Laboratory for Neutron Scattering and Imaging, Paul Scherrer Institut, 5232 Villigen, Switzerland
[2]Niels Bohr Institute, University of Copenhagen, Nørregade 10, 1165 Copenhagen, Denmark

*markus.strobl@psi.ch
$jacopo.valsecchi@psi.ch

**Abstract**
We demonstrate a simple single grating beam modulation technique, which enables the use of a highly intense neutron beam for phase imaging and thus spatially resolved structural correlation measurements in full analogy to quantum interference based methods. In contrast to interferometric approaches our method is intrinsically achromatic and provides unprecedented flexibility in the choice of experimental parameters. Utilizing merely a macroscopic absorption mask unparalleled length scales become accessible. The results presented, including application to a variety of materials, establish a paradigm shift in phase imaging.

State of the art imaging with visible light as well as with electrons comprises numerous contrast mechanisms, including phase and dark-field contrast schemes [1,2]. X-ray and neutron imaging, on the other hand, have only recently seen the development of an analogous variety of contrast modalities [3,4]. Today interferometric and non-interferometric techniques play an important role in phase imaging with all these types of radiation in providing contrast based on the real part of the refractive index, where conventional attenuation contrast fails.

Neutron interferometry has created significant impact in neutron imaging in particular when Talbot Lau grating interferometers enabled unprecedented high efficiency phase contrast imaging with neutrons early in the millennium [5]. Attempts to utilize interferometry and its outstanding sensitivity to quantum beam phase effects for spatially resolved measurements have been reported earlier and in particular with single crystal Mach-Zehnder interferometers [6]. However, their coherence and stability requirements as well as spatial restrictions hindered practical applications. Also attempted non-interferometric techniques suffered from high coherence requirements [7,8]. The ability of the grating interferometer to operate at relatively low coherence requirements overcame these restrictions. Thus, it unlocked the potential of the highly penetrating quantum beam of neutrons to visualize through phase contrast e.g. magnetic structures. Studies of vortex structures, magnetic domains and domain wall kinetics in the bulk of superconductors and ferromagnets which were not amenable to any other technique have been reported [9-14]. In addition, the high sensitivity of the apparatus was found to qualify it an excellent tool for the investigation of micro-structural features based on coherent scattering length density variations in bulk materials and objects [15-18]. In contrast to conventional instrumentation, the combination of microstructural sensitivity with macroscopic real space resolution enables the study of heterogeneous structures and processes in representative volumes [19-21]. Due to analogies to dark-field microscopy, this phase imaging technique is in x-ray and neutron imaging referred to as dark-field contrast imaging [15,22].

Today grating interferometers for imaging are available at many leading instruments at neutron sources around the world [5,15,23-26]. Numerous successful studies [9-21] triggered continuous developments with regards to varying implementations of analogue techniques [10,25-30]. Advances targeted mainly efficiency, complexity and assessable correlation length scales. Key progress was established in particular by the quantitative interpretation of dark-field contrast providing correlation length information [17,18]. The first quantitative characterization of microstructures was reported on the nano-scale already when a spin-echo interferometer was successfully introduced to neutron imaging [28]. It operates in full analogy to Talbot Lau grating interferometric imaging but builds on the quantum spin phase and interference of the two spin states of the neutron wave function. The advantage of the this method is access to the nano-scale, full remote control of the modulation as well as its suitability for the most efficient time-of-flight wavelength dispersive measurement approach with neutrons at pulsed spallation sources [28,29]. Disadvantages are the sophisticated and elaborate set-up and technological barriers to assess the micro-scale, without excessive sample to detector distances. This scale is however covered by the common Talbot Lau interferometry [18]. A more recent implementation, referred to as far-field multi-phase-grating interferometer [30], implies similar issues in the conventional range, but in addition faces efficiency issues due to a high general collimation requirement. An advantage is meant to be a lower wavelength dependence of the visibility achieved in the beam modulation.

Here we introduce a paradigm shift demonstrating that the interferometric nature of the applied techniques is not relevant but only the spatial modulation of the beam, no matter how it is achieved. In contrast to previous works underlining the virtues of quantum particle-wave interferometry we chose the most basic approach to create spatial beam modulation on a suitable length scale to retrievably encode small angle beam deviations superimposed to real space images [17]. We prove that we are thus able to record and analyse the corresponding multi-modal images providing attenuation contrast, differential phase and dark-field contrast in full analogy to the discussed interferometric techniques. In addition, a close analogy to non-interferometric phase imaging through structured illumination in light microscopy [31,32] can be found in this approach.

The set-up is established by merely adding an attenuation grating in a conventional pinhole collimated imaging instrument (Fig. 1). The image of the grating will be an accordingly spatially modulated intensity profile. Due to the geometric blur the modulation image will feature a sinusoidal intensity distribution, comparable to these of interferometric methods. The achievable visibility $V=(I_{max}-I_{min})/(I_{max}+I_{min})$ depends, for a perfect absorption grating, only on the resolution capability of the set-up and is fully independent of the wavelength, hence fully achromatic in contrast to any approach presented earlier (Fig.1b, c). This modulation of the beam which is simply a projection image of an absorption grating will be used in analogy to all interference based beam modulations before to provide quantifiable spatially resolved measurements of attenuation, phase and scattered wave interference. The principle of the latter is schematically depicted in Fig. 1a.

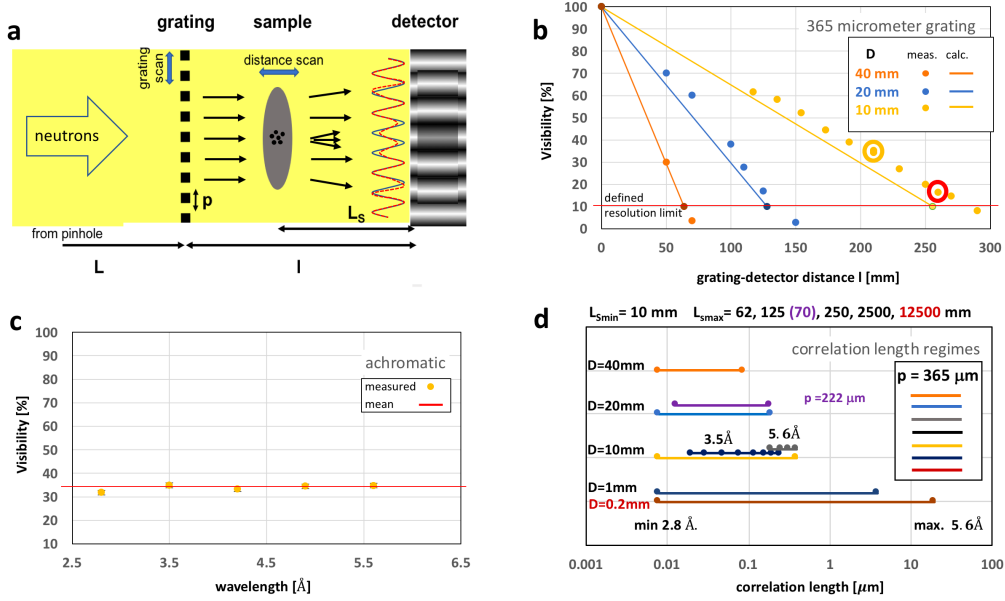

**FIG. 1 Set-up and parameters** (a) sketch of the basic set-up adding a simple attenuation grating at a distance L from the pinhole and upstream of a variable sample position in a conventional neutron imaging instrument. The principles of differential phase and scattering detection (dark-field contrast) are depicted. The blue curve represents the undisturbed pattern and red is the refracted (sample sides) and scattered (sample center) pattern. (b) visibilities V achieved for L=7m pinhole to grating distance for different standard pinhole sizes D = 40, 20 and 10 mm and the utilized absorption grating period 365 μm. The red line at 10% marks the theoretical resolution limit and the lines provide a conservative visibility estimate based on the calculated geometric blur d=l/(L/D). The point marked by a red circle indicates the parameters used for the presented study. (c) measured proof of achromatic visibility with data taken at the point marked with yellow circle in (b). (d) achievable correlation lengths according to eq. (2)[17] for standard pinhole sizes available at the benchmark instrument ICON at PSI [33] and D=0.2mm like utilized for far-field interferometry [30]. Minima of indicated ranges refer to $L_S$ = 10 mm and a minimum wavelength of 2.8 Å, maxima to $L_S$ according to (b) and a wavelength of 5.6 Å. At D=20 mm a comparison with the use of a 222 μm period grating is provided and at D=10 mm the parameters utilized in the presented study are shown.

The sensitivity of a modulated beam measurement technique depends on the modulation period $p$ and the sample to detector distance $L_s$ (Fig. 1a) and can in the small angle approximation be expressed by the characteristic angle or the probed scattering vector length

$$\theta_c = p/2L_s \quad \text{and} \quad q_c = \pi p/\lambda L_s \,, \tag{1}$$

respectively. At this angle, or scattering vector, intensity is shifted from the maxima exactly into the intensity depleted minima – the dark-field - of the modulation pattern. This in turn for the probed length scale equals [17]

$$\xi = \lambda L_s/p \quad . \tag{2}$$

For the intrinsically achromatic technique the wavelength λ is a free parameter (Fig. 1c), but it is limited by the typically used spectra in the thermal and cold energy range. Both other parameters, $L_s$ and $p$, are limited by the spatial resolution ability of the set-up (Fig. 1b). A larger distance and a smaller period provide access to larger structures (Fig. 1d). A typical Talbot Lau interferometer provides modulation conditions enabling to resolve micrometer sized structures with micrometer modulation periods. These require analyser gratings to be resolved. Here we consider periods of some 100 micrometer which can be resolved spatially up to some 100 mm distances by the detector directly. A workable compromise between visibility and resolvable range has to be found (Fig. 1b & d). Because the spatial resolution and hence the visibility at a certain grating to detector distance in a conventional pinhole collimated set-up is indirect proportional to the available flux density, also efficiency considerations have to be evaluated carefully. Theoretical approximations as well as measurements confirm the ability to create modulations with substantial visibility and compatible with outstanding correlation length ranges probed from the nanometer to micrometer scale with standard instrument settings (Fig. 1b & d). This applies despite the limited wavelength range of 2.8 to 5.6 Å accessible with the utilized instrumentation. Fig. 1 d, in addition, displays the unparalleled extensive range our technique covers when employing slits as reported to be applied for far field interferometry [30]. For our demonstration an intermediate regime with the potential to cover nearly two orders of magnitude in correlation lengths has been chosen while analyzed data are restricted to correlation lengths ranging from about 10 to 300 nm. This range clearly complements that of conventional Talbot Lau interferometry.

The measurements were performed with a 10 mm pinhole, common for conventional high resolution imaging measurements and leading to a collimation ratio $L/D$ of about 700. The chosen $L/D$ and grating to detector distance $l$ of 0.26 m enabled a visibility of 18% (Fig. 1b) comparable to such utilized in neutron Talbot Lau interferometry. The grating consisted of 20 µm high Gd lines on a quartz wafer with a duty cycle of 50% and a period of 365 µm. The detector used was a common combination of a 20 µm thick gadolinium oxysulfide (Gadox) scintillator screen and a commercial CCD camera (Andor, iKon-L) with a 100 mm Zeiss photographic lens system. The corresponding field of view (FoV) was 70 x 70 mm$^2$ with an effective pixel size of 35 µm and an intrinsic spatial resolution of around 70 µm. The grating had an effective size of 50 x 50 mm$^2$ and hence did not cover the full FoV of the detection system. The exposure time per image was 3 times 120 sec irrespective of the wavelength used.

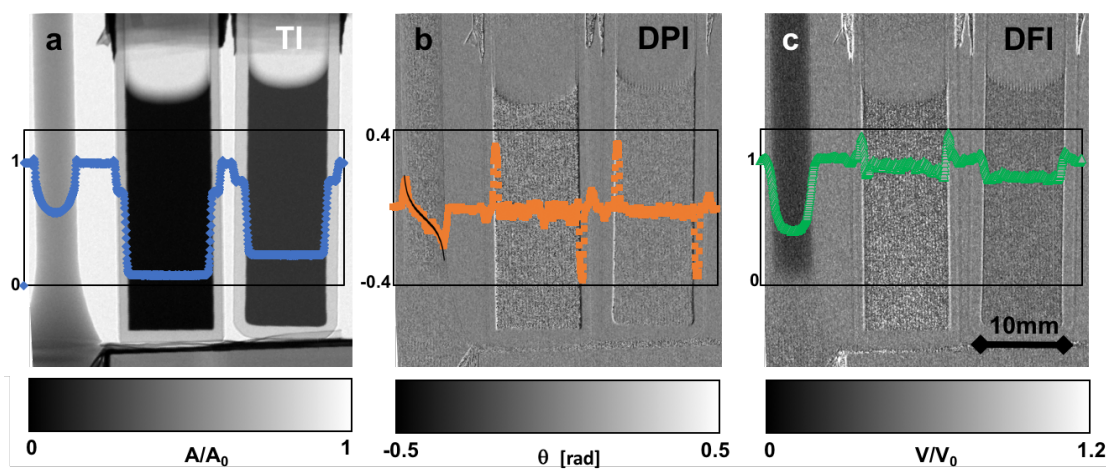

*Fig. 2 Three imaging modalities* are measured simultaneously and in analogy to grating interferometric imaging [9]; the images display a measurement of three samples, a 304L steel cylinder

and two quartz glass cuvettes containing aqueous solutions; (a) the conventional transmission image (TI) corresponds to the ratio between transmitted and incident beam intensities $A/A_0$ (eq. 3); (b) the differential phase image (DPI) is established by the modulation phase difference between open beam and sample θ (eq. 3), the line profile contains a theoretical calculation for the steel cylinder (black curve)(eq. 4); (c) the dark-field image (DFI) reflects the relative visibility loss through scattering from the sample $V/V_0=BA_0/(B_0A)$ (eq. 3);

A macroscopic beam modulation superimposed to an image enables different routes of measurement and analyses strategies for the contrast modalities [15,28,34,35]. While in principle it is possible to fully analyse such images for all modalities from a single shot [28,35], here, again in analogy to Talbot Lau interferometry, a grating scan has been performed with 11 steps over one period. This enables conventional pixel-wise extraction of all three contrast parameters: transmission $A$, differential phase parameter θ and visibility $V=B/A$ constituting the measured image as

$$I_{i,j} = A_{i,j} + B_{i,j} \sin(C_{i,j}X + \theta_{i,j}) \tag{3}$$

where i,j are the pixel indices, C is the phase of the open beam modulation and X is the grating scan parameter.

For an individual measurement of a set of samples the images corresponding to the three parameters: transmission image (TI), differential phase image (DPI) and dark-field image (DFI) are depicted in Fig. 2. The samples in this measurement are a cylindric tensile test dogbone sample of 304L steel and two aqueous solutions in quartz glass cuvettes placed in the beam. While the attenuation by the water in the 5mm thick cuvettes is substantial as can be seen in the attenuation contrast transmission image (TI) it affects the statistics in the corresponding areas in the differential phase image (DPI) and in the dark-field image (DFI) significantly. Nevertheless, all three contrast modalities display the typical corresponding features with respect to the samples, which are well-known from corresponding interferometric imaging approaches. Due to the good spatial resolution capability of the set-up and the significant sample to detector distance even the signature of far field phase contrast [7] is visible upon careful inspection in the attenuation contrast image (TI, Fig. 2a). The differential phase contrast image depicts the influence of the distorted neutron wave-front on the spatial modulation phase θ (Fig. 1a). The respective local neutron wave phase shift can be extracted from this according to the relation [5]

$$\theta = \frac{\lambda L_S}{p}\frac{\partial \Theta}{\partial x} \tag{4}$$

where $\partial\Theta/\partial x$ is the gradient of the neutron wave-front perpendicular to the modulation. The phase can be retrieved straightforwardly from corresponding integration [5]. For the phase profile of the cylindric steel sample we compare the measurement to the according calculation of the differential phase in Fig. 2b, underlining good quantitative agreement within the spatial resolution limit.

Systematic quantitative reference measurements have been performed on a second set of samples. The study focusses on quantitative dark-field imaging [15,17-21] which has established as the dominant application of interferometeric modulated beam imaging in material research. The first reference samples were two commercially available fractal powders, namely Sipernat-350 and Sipernat-310 with characteristic particle sizes of 4 and 8.5 micrometers, respectively [36]. Sipernat-310 is a powder of silica featuring a large surface area of 700 m$^2$/g. The large surface area together with the micrometer sized particle characteristics makes Sipernat-310 a well-suited material to investigate basic characteristics of cohesive powders. It has recently been used in a study observing with spatial resolution the heterogeneous breakdown of the fractal microstructure [21]. This study provided the required reference data and model fit from more conventional spin-echo small angle neutron scattering measurements with no direct spatial resolution.

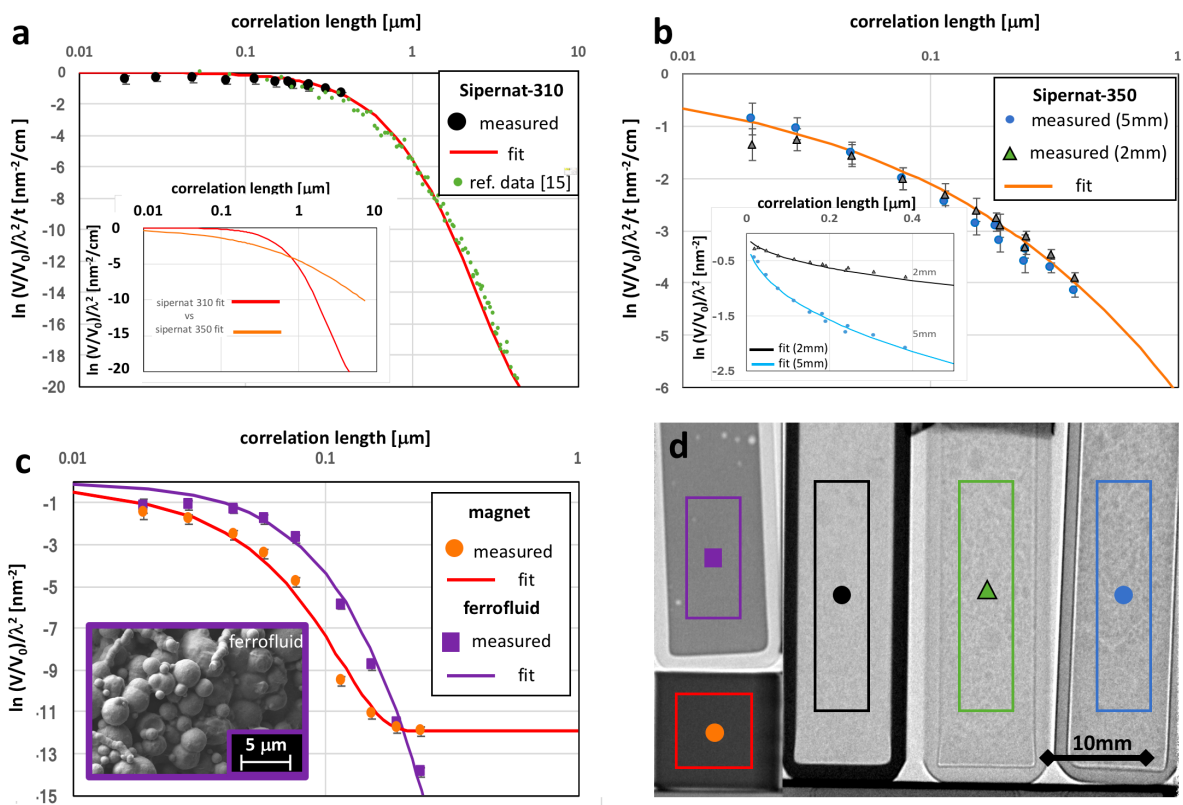

*FIG. 3 Mapping scattering length density correlations* Quantitative dark-field imaging performed through variation of wavelength and sample to detector distance at set-up parameters as indicated in Fig. 1b & d and applied to a number of samples as shown in the TI in (d); (a) results of measurement on a silica powder (Sipernat-310 [36]) combined with SESANS data and a model fit from literature [21]; the inset compares the applied model to the model used for the related powder sample presented in (b). Results of correlation length scan on two thicknesses of Sipernat-350 [36] powder with model fit (red line); insert displays the model applied separately to the results for the two different sample thicknesses before normalization. (c) data and fit for ferrofluid sample (violet squares) and a porous solid ferromagnet cube (red circles) and corresponding model fits. (d) an attenuation contrast TI indicating samples with symbols and colors used in the corresponding plots. Note, that where the error bars are not visible they are lying within the size of the used symbols.

Both powders were measured contained in quartz glass cuvettes with sample thicknesses of 5 mm for both Sipernat-310 and -350 and additionally a 2mm thick sample of Sipernat-350. Distance scans were performed at wavelengths of 3.5 Å and 5.6 Å from $L_s$ 20 to 250 mm and

120 to 250 mm, respectively (compare Fig. 1b). The measured visibilities from grating scans at each of these settings were analysed pixel-wise using our standard Talbot Lau data reduction software [37]. The normalized visibilities can be written as [17]

$$\frac{V(\xi)}{V_0} = e^{\Sigma t(G(\xi)-1)} \qquad (5)$$

where $V_0$ is the visibility without sample, $\Sigma$ is the total small angle scattering cross section and $G(\xi)$ is the projected real space correlation function of the microscopic sample structures [17]. The data is further reduced pixel-wise by computing the logarithm and dividing by the sample thickness $t$ and the wavelength square. The total scattering probability $\Sigma t$ depends on $\lambda^2$, which needs to be normalized in order to combine scans at different wavelengths and to achieve results independent of the utilized beam. Subsequently the data is modelled with specific projected real space correlation functions $G(\xi)$ representative of the microscopic structure investigated. A fitting procedure finally enables to retrieve the respective structural parameters.

According to Ref. [21] Sipernat-310 is best modelled with a randomly distributed two-phase medium [38] corresponding to

$$G(\xi) = \frac{\xi}{a} K_1\left(\frac{\xi}{a}\right) \qquad (6)$$

where $K_1$ is the modified Bessel function of second kind and first order and $a$ represents the characteristic structure size in the two-phase medium. The model and parameters found through small angle scattering measurements in Ref.[21], where $a$ = 1.56 μm and $\Sigma$ = 0.45, fit the single grating data as depicted in Fig. 3a. This model, however, did not fit the data of Sipernat-350, the second powder sample (compare inset Fig. 3a). This correlates with the fact that the powders are significantly different in their specific structure and surface area, which for Sipernat-350 is 55 m$^2$/g, compared to 700 m$^2$/g of Sipernat-310 [36]. The two data sets from two different sample thicknesses of Sipernat-350, Fig. 3b inset, could both be fitted with the same model

$$G(\xi) = e^{[-(\xi/a)^\alpha]} \qquad . \qquad (7)$$

This model describes a simplified limited fractal taking into account an exponential cut-off [39]. The two data sets after normalization with their respective thicknesses coincide well (Fig. 3b). The fit yields a characteristic length scale $a$ = 7 μm and $\alpha$ = 0.5, with $\alpha$ being related to the structure of the phase boundary. Here $\alpha$ ranges in the lower domain of 0< $\alpha$ <1 which corresponds to open and branched distributions with high specific surfaces.

Additional data on well characterized samples returned the porosity on the probed length scale of a ferrite magnet produced through sintering as well as the poly-disperse structure size distribution in a ferrofluid (Fig. 3c). The cubic ferrite sample of 10 mm side length consisting of $Fe_2O_3$ (86%) and $SrCO_3$ (14%) is revealed to have significant porosity in the probed size range constituting about 1% of the volume. The pores are characterized with a poly-disperse size distribution with predominant sizes of 20 and 210 nm diameter. Hard sphere models [17] were used to approximate the pore structure, and remaining deviations can be assumed to be due to deviation of average pore shape from the spherical model. The volume fraction of the smaller pores is found to be more than a factor two lower than the one of the dominating 200 nm pores constituting a volume fraction of 0.6%. In addition, microscopy reveals, that pores can also be found on the scale of several micrometers, which is, however, beyond the probed length scale and does not influence the presented results. The ferrofluid can be described by a random two-phase medium (Eq. 6) consisting of a polydisperse particle distribution (inset Fig. 3c) suspended in a liquid. Analyses provides a characteristic size parameter of $a$ = 250 nm, corresponding to the smallest particles found in the liquid. In addition, the largest length scale derived from model fitting coincides with the largest particle dimension in the ferrofluid of about 3 $\mu$m (Fig. 3c). The deviations of the measured points from the fit at the lowest probed correlation lengths are understood to be due to incoherent scattering from the hydrogenous liquid phase.

We conclude, that our most simple approach of a single attenuation grating with sub-millimeter period that requires no specific and sensitive alignment, does overcome stability requirements, design and fabrication issues related to fine structures as well as other limitations of interferometers. In particular the intrinsically achromatic nature of our technique enables straightforwardly multi-wavelength and polychromatic studies without the corresponding drawbacks of interferometry. In contrast, it provides the capability for efficient, simple, flexible and quantitative multi-modal imaging and especially phase imaging in the form of differential phase and dark-field contrast. In addition, the method holds the potential to straightforwardly cover an unprecedented correlation length range. The capability to extend the range reported for Talbot Lau interferometers [17-21] by more than an order of magnitude into the nanometer range has been demonstrated. Flux densities available even at medium flux sources allow to further increase collimation ratios within standard imaging settings (Fig. 1). This enables to substantially increase the total covered correlation length scale range from the nanometer regime into and beyond the typically assessed micrometer range of neutron Talbot Lau interferometry. In particular a comparison with recent approaches of far-field interferometers [30] (compare Fig. 1d) suggests that similar collimation conditions lead to significantly superior performance parameters. This implies that our much simpler and in contrast non-interferometric and truly achromatic set-up can provide higher efficiency with at least the same accuracy and range of experiments without the need for careful alignment, stability and interferometry per se. Thus, the introduced approach constitutes a paradigm shift for corresponding measurements and methods.

Furthermore, especially the wider spectral range accessible continuously at high flux spallation neutron sources generates substantial benefit from the achromatic nature of the set-up for large range high flux measurements including kinetic studies. Instruments like the imaging beamline ODIN under construction at the European Spallation Source [40] cover wavelength ranges of up to one order of magnitude. This implies that a correlation length range of the same order of magnitude can be probed simultaneously, enabling the spatially resolved observation of microstructural changes not only in time-dependent but non-uniform environments such as e.g. in shear fields, flow, inhomogeneous temperature, pressure and magnetic fields in full field observations. This paves the way to foster entirely new analytical capabilities for complex inhomogeneous and non-equilibrium structural states in materials either opaque to or not providing sufficient contrast with other types of radiation.

An extension to 2-dimensional phase contrast sensitivity and resolution is similarly straight forward through the utilization of 2D absorption patterns in contrast to recent elaborate approaches with interferometry [27,41]. This enables to additionally study micro-structural anisotropies without the need of several sample scans.

Extending this approach to x-rays is expected to be similarly rewarding because it is suited to profit from the superior phase space densities available from x-ray sources. These allow projecting small periods at significant distances, thus covering an outstanding correlation length scale range.